\documentclass[twocolumn]{aastex61} % STANDARD STYLE TO SUBMIT

\received{XX 1, 2017}
\revised{XX 1, 2017}
\accepted{\today}
\submitjournal{ApJ}

\shorttitle{A novel method to detect and measure the ages of star clusters}
\shortauthors{Bitsakis et al.}
\begin{document}

\title{A novel method to automatically detect and measure the ages of star clusters in nearby galaxies: Application to the Large Magellanic Cloud}

\correspondingauthor{Theodoros Bitsakis}
\email{t.bitsakis@crya.unam.mx}

\author{T. Bitsakis}
\affiliation{Instituto de Radioastronom\'ia y Astrof\'isica, Universidad Nacional Aut\'onoma de M\'exico, Morelia, 58089, Mexico}

\author{P. Bonfini}
\affiliation{Instituto de Radioastronom\'ia y Astrof\'isica, Universidad Nacional Aut\'onoma de M\'exico, Morelia, 58089, Mexico}

\author{R. A. Gonz\'alez-L\'opezlira}
\affiliation{Instituto de Radioastronom\'ia y Astrof\'isica, Universidad Nacional Aut\'onoma de M\'exico, Morelia, 58089, Mexico}
\affiliation{Helmholtz-Institut f\"ur Strahlen-und Kernphysik (HISKP), Universit\"at Bonn, Nussallee 14-16, D-53115 Bonn, Germany}
\affiliation{Argelander Institut f\"ur Astronomie, Universit\"at Bonn, Auf dem H\"ugel 71, D-53121 Bonn, Germany}

\author{V. H. Ram\'irez-Siordia}
\affiliation{Instituto de Radioastronom\'ia y Astrof\'isica, Universidad Nacional Aut\'onoma de M\'exico, Morelia, 58089, Mexico}

\author{G. Bruzual}
\affiliation{Instituto de Radioastronom\'ia y Astrof\'isica, Universidad Nacional Aut\'onoma de M\'exico, Morelia, 58089, Mexico}
 
\author{S. Charlot} 
\affiliation{Sorbonne Universit\'es, UPMC-CNRS, UMR7095, Institut d'Astrophysique de Paris, F-75014 Paris, France}

\author{G. Maravelias}
\affiliation{Instituto de F\'isica y Astronom\'ia, Universidad de Valpara\'iso, Valpara\'iso, Chile}
\affiliation{Astronomick\'y \'ustav, Akademie v\v{e}d \v{C}esk\'e republiky, Fri\v{c}ova 298, 251\,65 Ond\v{r}ejov, Czech Republic}

\author{D. Zaritsky}
\affiliation{Steward Observatory, University of Arizona, Tucson, AZ 85719, USA}



\begin{abstract}

We present our new, fully-automated method to detect and measure the ages of star clusters in nearby galaxies, where individual stars can be resolved. The method relies purely on statistical analysis of observations and Monte-Carlo simulations to define stellar overdensities in the data. It  decontaminates the cluster color-magnitude diagrams and, using a revised version of the Bayesian isochrone fitting code of Ram\'irez-Siordia et al., estimates the ages of the clusters. Comparisons of our estimates with those from other surveys show the superiority of our method to extract and measure the ages of star clusters, even in the most crowded fields. An application of our method is shown for the high-resolution, multi-band imaging of the Large Magellanic Cloud. We detect 4850 clusters in the 7 deg$^{2}$ we surveyed, 3451 of which have not been reported before. Our findings suggest multiple epochs of star cluster formation, with the most probable occurring $\sim$310 Myr ago. Several of these events are consistent with the epochs of the interactions among the Large and Small Magellanic Clouds, and the Galaxy, as predicted by N-body numerical simulations. Finally, the spatially resolved star cluster formation history may suggest an inside-out cluster formation scenario throughout the LMC, for the past 1 Gyr.

\end{abstract}




%%%%%%%%%%%%%%%%%%%%%%%%%%%%%%%%%%%%%%%%%%%%%%%%%%

%%%%%%%%%%%%%%%%% BODY OF PAPER %%%%%%%%%%%%%%%%%%

\section{Introduction}
Star clusters provide a unique tool to understand the star formation history of Local Group galaxies, given their well constrained distances and our ability to determine their ages with good precision. They can also provide significant information about the initial mass function (IMF), the distribution of star formation within a galaxy, and whether star formation occurs everywhere synchronously \citep{Maragkoudakis17}. 

The Large Magellanic Cloud (LMC) is not only one of the nearest galaxies, it is also an exceptional laboratory of galaxy evolution under the influence of strong gravitational interactions. The LMC and the Small Magellanic Cloud (SMC), constitute an interacting pair, which is also bound to the Galaxy. It is believed that the Magellanic Stream --- an 180 kpc long intergalactic filament that extends from the LMC to the SMC and through the Galactic south pole --- is a relic of strong tidal encounters between the aforementioned galaxies \citep[see][]{Yoshizawa03, Besla12}. 

Various authors have attempted to study the formation of star clusters in the LMC. \citet{Bica08} compiled a general catalog, using all the previously known sources, as well as their own findings based on visual inspection on Sky Survey plates. Their catalog contains 9305 extended objects in the Magellanic System (the LMC, the SMC and the stream); among these, about 3700 have been identified as clusters/associations in the LMC. Using the Bica et al.  catalog, \citet{Glatt10} estimated the cluster ages through visual fitting of their color-magnitude diagrams (CMDs) with isochrones generated by the Geneva and Padova codes \citep[][respectively]{Lejeune01,Girardi95}. They concluded there were various periods of intense star cluster formation over the last 1 Gyr, the two most prominent of which occurred 125 and 800 Myr ago. In addition, \citet{Baumgardt13} compiled a new catalog with the ages of 307 clusters by selecting the best age estimations from previous publications. Their data suggest a burst of cluster formation about 1 Gyr ago, with little evidence for the existence of similar activity afterwards. More recently, \citet{Nayak16} presented a semi-automated method to estimate age and reddening of 1072 star clusters in the LMC. They found at least one major cluster formation event 125 Myr ago. There seems, therefore, to be a consensus that the LMC experienced various episodes of star formation, plausibly compatible --- given the uncertainties --- with the results of N-body simulations on the tidal interactions between the Magellanic Clouds and/or the Galaxy. Recently, such simulations which made use of the latest proper motion observations, suggested that the Magellanic Clouds are probably in their first passage about the Galaxy \citep[e.g.][]{Besla07, Besla12, Kallivayalil13}.

Despite all this progress, the aforementioned studies exploring the age distribution and properties of star clusters in the LMC did not only demand significant time and effort to be conducted, but they also introduced important biases (as described in Sections 3 and 4). \citet{Piatti12} have shown that, when not correctly addressing the field star contamination, one can detect more false associations than real star clusters --- owing to stochastic fluctuations. In this work, we introduce the first, fully automated method able to detect and determine the ages of star clusters in nearby galaxies. The detection is performed using a novel spatial clustering procedure that is described analytically in Section 3, while in Section 4 we present the age identification algorithm. Since this method is fully computer-based, it avoids  important biases that may be introduced by the limitations of solely visual identification; it is also more suited to the use of clustering thresholds and multi-wavelength data. Using the results of our algorithms, in Section 5 we present the LMC star cluster properties, as well as their age and spatial distributions. Finally, we briefly give our summary and conclusions in Section 6.   

Throughout this work we assume a distance modulus for the LMC of 18.50 mag \citep{Walker11}.

% ################################################
\section{The data}
\subsection{Imaging}

\begin{figure*}
\begin{center}
\includegraphics[scale=0.36]{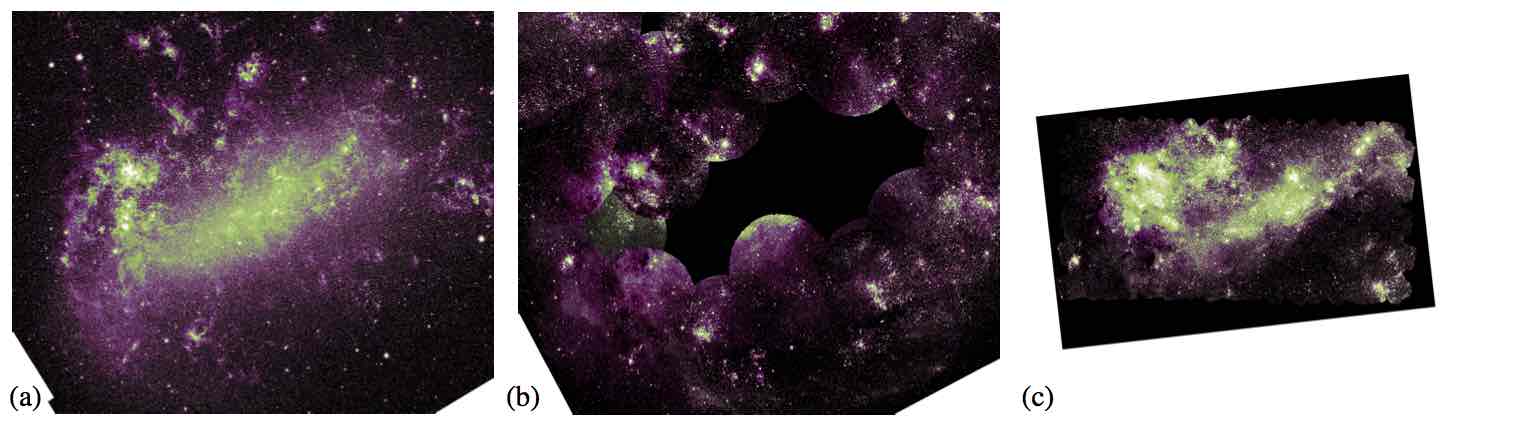}
\caption{An example of the multi-band data we used in our work: (a) a Spitzer/IRAC 3.6$\micron$ mosaic of the central region of the LMC \citep{Meixner06}, (b) the GALEX/$NUV$ mosaic without the central bar \citep{Simons14}, and (c) the central region observed by Swift/UVOT \citep{Siegel14}.}
\label{fig1}
\end{center}
\end{figure*}

The archival data used in this work were acquired from several diverse large surveys which mapped the Magellanic Clouds at various bands (some examples are shown in Fig.~\ref{fig1}). Starting from shorter wavelengths, \citet{Simons14} composed a mosaic using archival data from the  Galaxy Evolution Explorer \citep[GALEX;][]{Martin05} at the near-ultraviolet ($NUV$) band ($\lambda_{\rm eff}$=2275\AA). The mosaic covers an area of 15 deg$^{2}$ on the LMC. All exposures of the same field were co-added to improve the signal-to-noise ratio. The median exposure time was 733 seconds, and the 5$\sigma$ depth varied between 20.8 and 22.7 mags. Unfortunately, this mosaic does not cover the central $\sim$3$\times$1 deg$^{2}$ of the LMC (the bar-region), due to detector limitations. This area was later observed by the Swift Ultraviolet-Optical Telescope (UVOT) Magellanic Clouds Survey \citep[SUMAC;][]{Siegel14}, which covered an area of $\sim$4$\times$2 deg$^{2}$ around the bar-region, with typical exposures of 3000 sec in all three $NUV$ filters of this instrument ($UVW1$, $UVW2$, and $UVM2$). 

The optical data used here are from the Magellanic Cloud Photometric Survey \citep[MCPS;][]{Zaritsky04}. These authors observed the central 64 deg$^{2}$ of the LMC with 3.8-5.2 minute exposures at the Johnson $U$, $B$, $V$, and Gunn $i$ filters of the Las Campanas Swope Telescope. Typical seeing was 1.5 arcsecond, with limiting magnitudes that varied, depending on the filter, between 21.5 mag for $U$ and 23.0 mag for $i$. 

\citet{Meixner06} performed a uniform and unbiased imaging survey of the LMC (called Surveying the Agents of a Galaxy's Evolution, or SAGE), covering the central 7 deg$^{2}$ with both the Infrared Array Camera (IRAC) \citep{Fazio04} and the Multiband Imaging Photometer \citep[MIPS;][]{Rieke04} on-board the Spitzer Space Telescope. The SAGE survey produced mosaics at 3.6, 4.5, 5.8, and 8.0$\micron$ for IRAC, and at 24, 70, and 160$\micron$ for MIPS. The exposure times varied between 43-60 seconds, depending on the band, and added up to a total of 291 hours for IRAC and 217 hours for MIPS. Our current analysis has been performed on the area covered by SAGE, and we cropped accordingly the mosaics of GALEX, SUMAC, and MCPS surveys. 

\subsection{Photometric catalog and extinction corrections}
Using DAOPHOT II \citep{Stetson87}, \citet{Zaritsky04} elaborated a photometric catalog which contains 24.5 million sources in the area covered by MCPS. They also estimated the line-of-sight extinctions of the stars in their catalog to produce an extinction map of the LMC. To that end, they  compared the observed stellar colors with those derived by the stellar photospheric models of \citet{Lejeune97}. Thus, they measured the effective temperature ($T_{\rm eff}$) and the extinction ($A_{V}$) along the line of sight to each star, adopting a standard Galactic extinction curve. They produced two $A_{V}$ maps, one for hot (12000 K $<$ $T_{\rm eff}$ $\le$ 45000 K) and one for cool (5500 K $<$ $T_{\rm eff}$ $\le$ 6500 K) stars. 

Using these extinction maps, we correct the observed colors of the cluster candidate and field comparison stars, after separating them into two categories depending on their color (and thus their $T_{\rm eff}$). Stars having ($B-V$)$\le$0.20, which corresponds to A5 or earlier type stars with $T_{\rm eff}\ge$7800 K, are classified as hot and corrected with the hot star map, whereas for stars with ($B-V$)$>$0.20 we use the cool star map. We also adopt $R_{V}$=3.1 and relative extinctions $A_{\lambda}$/$A_{V}$ from \citet{Schlegel98}. 

% ################################################
\section{The cluster detection methodology}
\subsection{Selecting the appropriate extraction algorithm}
Detecting spatial clustering of stars in the Galaxy as well as in nearby galaxies, where individual stars can be observed, is a challenging task. Since star clusters are not found in isolation and many times are projected over very crowded fields (e.g., on the central bar of the LMC), it is nearly impossible for a visual search to distinguish cluster members from field stars. An automated method that is based on statistical analysis and Monte-Carlo simulations is arguably more prone to succeed. \citet{Schemja10} presented a comparison between four different cluster detection algorithms, which we briefly describe in the following: (i) the star counts, (ii) the nearest neighbor, (iii) the Voronoi tessellation, and (iv) the separation of the minimum spanning tree. The star counts method simply counts stars located in a region-of-interest and detects overdensities above some local background threshold that is defined by the user. The nearest neighbor method estimates the local source density by measuring the distance of each object to its $n$-th nearest neighbor. A Voronoi tessellation partitions an ($x$,$y$) plane with $n$ points into $n$ polygons, such that each polygon contains only one point, and then defines the local source density as the reciprocal of the area of the polygon around this point. Finally, the separation of the minimum spanning tree traces a unique set of lines (called edges), connecting a given set of vertices without closed loops, such that the sum of the edge lengths is minimum; by applying the desired length threshold, star clusters can be defined. As \citet{Schemja10} explains, he carried out Monte-Carlo simulations of synthetic star clusters using simple power-law, Gaussian or King spatial stellar distributions. Then he examined the performance of the aforementioned algorithms and concluded that, while distinct centrally concentrated clusters are detected by all methods, those with low overdensity or highly hierarchical structure are only reliably detected by methods with inherent smoothing, i.e., the star counts and nearest neighbor algorithms. Based on these considerations, and taking into account that the latter requires significantly larger computational time ($\times$200 longer), the star counts method was selected as the optimal approach for our analysis. 

\subsection{Implementing the detection sequence}

\begin{figure}
\begin{center}
\includegraphics[scale=0.22]{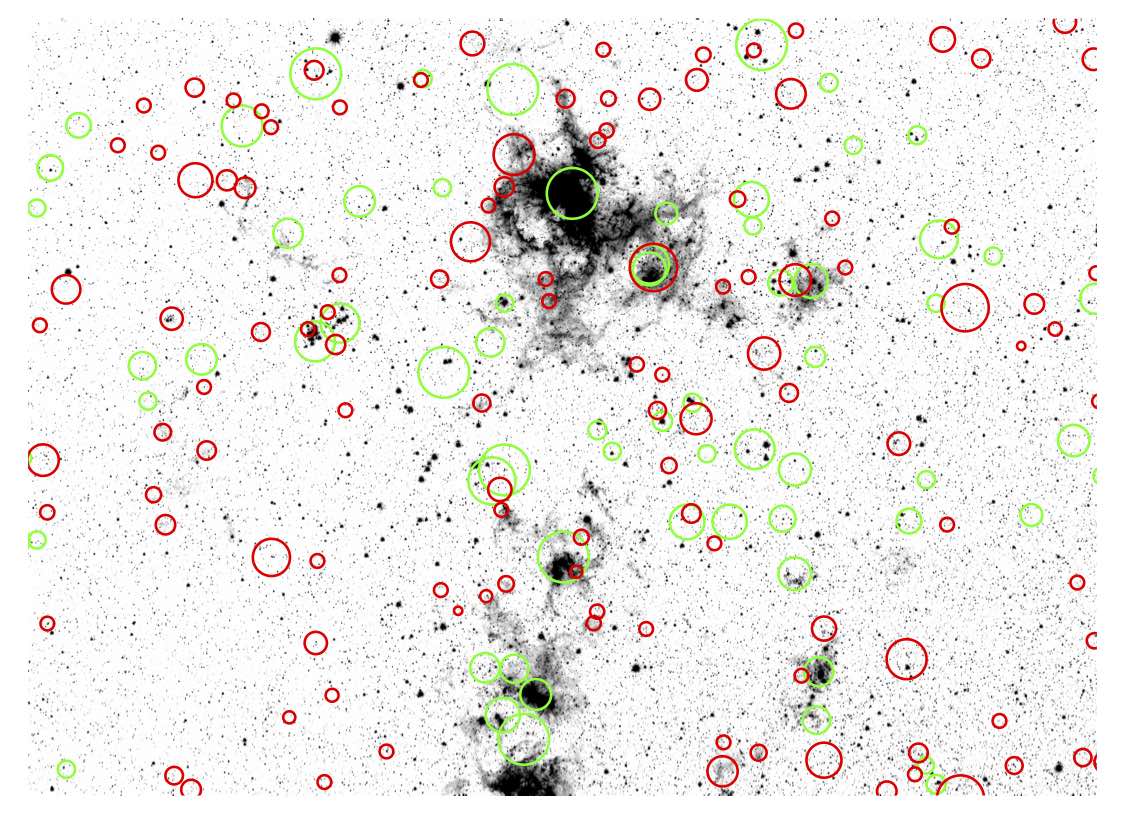}
\caption{Detail of the candidate cluster detection process near the 30 Doradus region of the LMC. {\it Green circles} correspond to detections made on the GALEX/$NUV$ data, while {\it red circles} correspond to those made on Spitzer/IRAC data. The background image is the Spitzer/IRAC 3.6$\micron$ from \citet{Meixner06}. }
\label{fig2}
\end{center}
\end{figure}

\begin{table*}
\begin{minipage}{120mm}
\begin{center}
\caption{Recovery rates from the results of Monte-Carlo simulations.}
\label{tab_recovery}
\begin{tabular}{lccc}
\hline 
\hline
 Telescope/ & Background density  & Artificial cluster & False association \\
 Filter &  pixel-stars deg$^{-2}$ & recovery rate (\%)$^{a}$ & detections (\%)$^{b}$ \\
\hline
Spitzer $IRAC1$ & 10$^{5}$ (high)$^{c}$ & 97 & 92 \\
Spitzer $IRAC1$ & 10$^{4}$ (low) & 97 & 10 \\
GALEX $NUV$ &  2$\cdot$10$^{4}$ (high) & 100 & 3 \\
GALEX $NUV$ &  10$^{3}$ (low) & 95 & 13 \\
Swift $UVW1$ &  5$\cdot$10$^{4}$ (high) & 70 & 7 \\ 
\hline
\end{tabular}
\end{center}
{\bf Notes. }\\
$^{a}$ Fraction of artificial clusters recovered by the code.\\ 
$^{b}$ Fraction of false associations with respect to the total number of artificial clusters.
$^{c}$ The LMC bar is considered a high background density region; low background density regions are more than 3.5 deg away from the bar. \\
\end{minipage}
\end{table*}% 

Images of reasonable resolution and depth (i.e., allowing to observe and resolve individual stars) are crucial for the identification of the candidate clusters. After detecting the positions of all the stars in an image through a source extraction code \citep[e.g., SExtractor;][]{Bertin96}, we create a pixel-map, where each star (or extended object) is represented by a single pixel. This is done to remove the effects of seeing and extended sources (e.g., galaxies) from the subsequent analysis. All these pixel-stars are assigned the same arbitrary flux value, since their spatial density, and not the photometric properties, is the important parameter for the cluster detection sequence. Once the conversion of the observed image to a pixel-map is done, the star cluster detection code is applied. The main virtue of the code is that it only requires to adjust two parameters: the size of the region-of-interest and the detection threshold. The region-of-interest is a box where the density of stars is counted and compared to the local background density, which is estimated over a much larger box, whose size is not critical as long as it samples well the local background density. After accurate testing, we observed that the ideal size of the region-of-interest is approximately that of a cluster at the distance of the LMC \citep[$\sim$60 arcsec in diameter;][]{Nayak16}, since it minimizes the detection of false associations. 

A more critical parameter to set is the detection threshold ($\Sigma_{\rm det}$). As described in \citet{Schemja10}, large values of $\Sigma_{\rm det}$ will be able to detect real overdensities only in low background density regions. On the other hand, low $\Sigma_{\rm det}$ values may be able to extract clusters in high background density regions (e.g., the LMC bar), at the cost of detecting many false associations (results of random projections) in the lower density regions, such as the outskirts of the galaxy. For this reason our code is using a variable $\Sigma_{\rm det}$ value that changes as a function of the local background density. To define the relation between $\Sigma_{\rm det}$ and the background density, we performed Monte-Carlo simulations with artificial clusters having both Gaussian as well as uniform overdensity profiles (accounting for both compact and diffuse clusters), projected over various background values. In Table~\ref{tab_recovery} we list the background densities (both high and low), and the rates of artificial cluster recovery and false association detection, in all the different filters we used during the extraction sequence. Our variable $\Sigma_{\rm det}$ is able to maintain a constant detection rate.

Using this $\Sigma_{\rm det}$ calibration, we run the detection code  to identify star clusters. At this step, pixel-stars which are considered as cluster members are recorded, while pixel-stars deemed as background ``counts'' are excluded from further analysis. Once we are left with the ``good'' pixel-map (i.e., where each pixel-star is a bona-fide member of a cluster), we need to determine the cluster center and radius. To do so, we use again SExtractor, which is able to detect flux emitted by coherent pixel groups, and provide their centroid and size. We therefore need to ``fool'' SExtractor into interpreting our good pixel-map as a flux map. We achieve this by smoothing the good pixel-map with a large kernel (similar to that of the point spread function), to redistribute the arbitrary flux in each pixel-star over the neighboring pixels. This will give a flux-like image over which we can run SExtractor using its intuitive setup. Since the flux-like image is generated from the pixel-map that contains only cluster members, it does not suffer from any ambiguity related to the background. In practice, we recreated the original image, but background subtracted (where ``background'' in this context means  ``field stars''). We have to stress here that, due to stellar resolution incompleteness in the most crowded regions of massive clusters (representing $<$1\% of our sample), there might be an overestimation of the final cluster radius of the order of 10-20\%. %Yet this will not further affect our candidate cluster stellar analysis, since it is properly treated by our code at a later stage (see Sect. 4.1). 

To maximize the detection of as many clusters as possible (both young and old), we apply our detection sequence on images at different wavelengths (see Fig.~\ref{fig2}). For this purpose we have used imaging from GALEX and Swift in the UV (that probes the hot massive stars), and the Spitzer IRAC, which is dominated by old stellar populations and low mass stars. Our catalog contains 5459 cluster candidates in an area covering the central 7 deg$^{2}$ of the LMC. For each cluster center and radius (defined by the SExtractor), we find all the stars within by cross-correlating with the extinction corrected MCPS catalog (see Section 2.2). This procedure has produced a photometric catalog of 1.9 million stars located in our candidate clusters. 

% ################################################
\section{Determination of the star clusters' ages}
The determination of the age of a star cluster is a rather challenging process and can be performed using various techniques. One can fit the observed color-magnitude diagram (CMD) of a cluster with sets of theoretical isochrones --- this method is arguably the most commonly used --- or compare its observed integrated colors (or spectrum) with those derived by theoretical modeling.  Although in theory star clusters should present well-defined CMDs, in practice there are a number of factors that introduce significant noise and biases. Field star contamination, both from the LMC and the Galaxy; photometric limitations and incompleteness; metallicity gradients; multiple main sequence turn-offs (MSTOs); blue straggler and binary stars are some major complications. 

Previous attempts to determine the ages of the LMC star clusters relied mostly on visual identification techniques. For instance, \citet{Pietrzynski00}, \citet{Glatt10}, and \citet{Piatti15} determined the ages of their clusters through visually identification of the MSTO, after eliminating the field stars by examining the CMD of a region around each cluster. \citet{Popescu12} carried out a comparison of the integrated broadband photometry of each cluster with models, though without performing a field star decontamination. More recently, \citet{Nayak16} used a semi-automated method to estimate the ages. Although the final age determination was decided again by visually fitting the MSTO, a quantitative automated method was used to constrain the range of plausible ages and decontaminate the clusters. Finally, \citet{Asad16} used a code to fit the integrated spectra of star clusters with sets of theoretical models, albeit without removing the contribution of field stars from the spectra. In the next paragraphs we present a novel age determination method: it is a completely automated, Bayesian CMD fitting algorithm.  

\begin{figure}
\begin{center}
\includegraphics[scale=0.24]{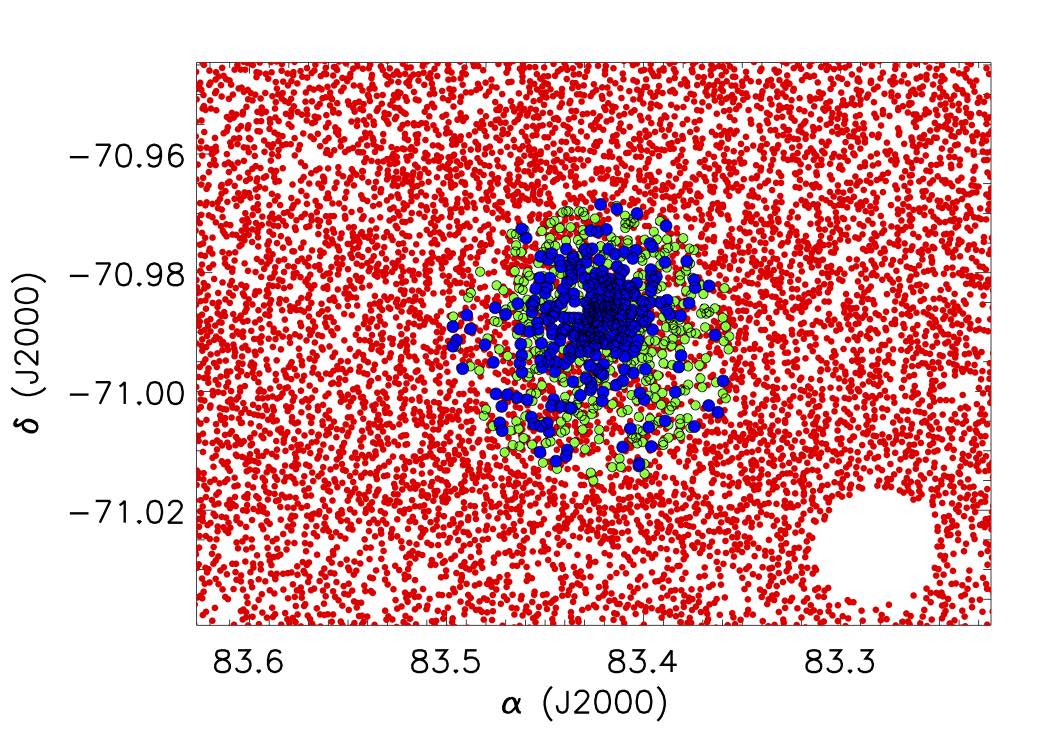}
\caption{Example of the decontamination procedure on the candidate cluster IR1-297. In this figure, stars are color-coded based on the membership probability ($p_{\rm memb}$) associated by our code. We see that stars with $p_{\rm memb}\ge$ 0.90 ({\it blue circles}) are concentrated in the center of the cluster, surrounded by stars with 0.60 $<$$p_{\rm memb}\le$ 0.90 ({\it green circles}). On the other hand, while several field stars (with $p_{\rm memb}\le$ 0.60; shown in {\it red circles}) are projected on the cluster, they are mostly distributed away from its center. Notice the empty region in the bottom right corner; it corresponds to stars masked-out by the code, since they belong to a neighboring cluster.}
\label{fig_decont}
\end{center}
\end{figure}

\subsection{Decontamination of the CMDs}

A crucial step in defining a cluster CMD is the proper accounting of the field star contamination. Here we use a process similar to that described in \citet{Mighell96}. For each cluster candidate CMD, our code also produces a field star CMD, using all the stars contained in a box of 0.16 deg$^{2}$ around the cluster center but excluding any stars located $\le$0.05 deg from the cluster. This is done to sample well the surrounding field star CMD without including any unaccounted cluster stars located outside the cluster radius. The CMDs are then binned along both axes using the color and magnitude uncertainties --- in our case we used $\delta$(color) = 0.5 mag and $\delta$(magnitude) = 1 mag. The code estimates the membership probability ($p_{\rm memb}$) of each star to belong to the given cluster, by considering the number of stars populating the corresponding bins in the cluster and  field CMDs. In particular, we adopted the following formula:
\begin{equation}
p_{\rm memb}=1-\frac{N_{\rm *,field}}{N_{\rm *,cluster+field}}\cdot\frac{A_{\rm cluster}}{A_{\rm field}} ,
\end{equation}
where $N_{\rm *,field}$ is the number of field stars (in the 0.16 deg$^{2}$ box), and $N_{\rm *,cluster+field}$ is the total number of stars contained within the cluster radius (i.e., a combination of field and cluster stars) in each color-magnitude bin. $A_{\rm field}$ and $A_{\rm cluster}$ are the sizes of the areas from which the field and cluster+field stars are extracted, respectively. Each cluster star candidate is hence assigned a $p_{\rm memb}$, which will later be used by the age determination code. An example of the results of this procedure for the candidate cluster IR1-297 is presented in Fig.~\ref{fig_decont}; stars with  $p_{\rm memb}\ge$ 0.90 are represented with blue circles, in green circles are those with 0.60 $<$ $p_{\rm memb}\le$ 0.90, while all other stars (having $p_{\rm memb}\le$ 0.60) are shown as red circles. Most stars in the central denser region of the cluster are assigned a higher membership probability, while green and red circles are mainly encountered in the periphery of the cluster. Therefore, our CMD decontamination is also consistent with the expected radial distribution of the clusters. It is also noteworthy that during this decontamination procedure we excluded from the field region all those stars that might belong to some neighboring clusters, hence keeping clean and unbiased the field star CMD (e.g., see empty space in the bottom right corner of Fig.~\ref{fig_decont}). Finally, the code discards as a false detection any cluster candidate with fewer than 20 stars with $p_{\rm memb}\ge$ 0.60. We imposed this additional step to eliminate any false associations detected during the cluster extraction process; the number of candidates found from star counts thus decreased from 5459 to 4850 ($\approx$ 11\% reduction). 

\begin{figure}
\begin{center}
\includegraphics[scale=0.24]{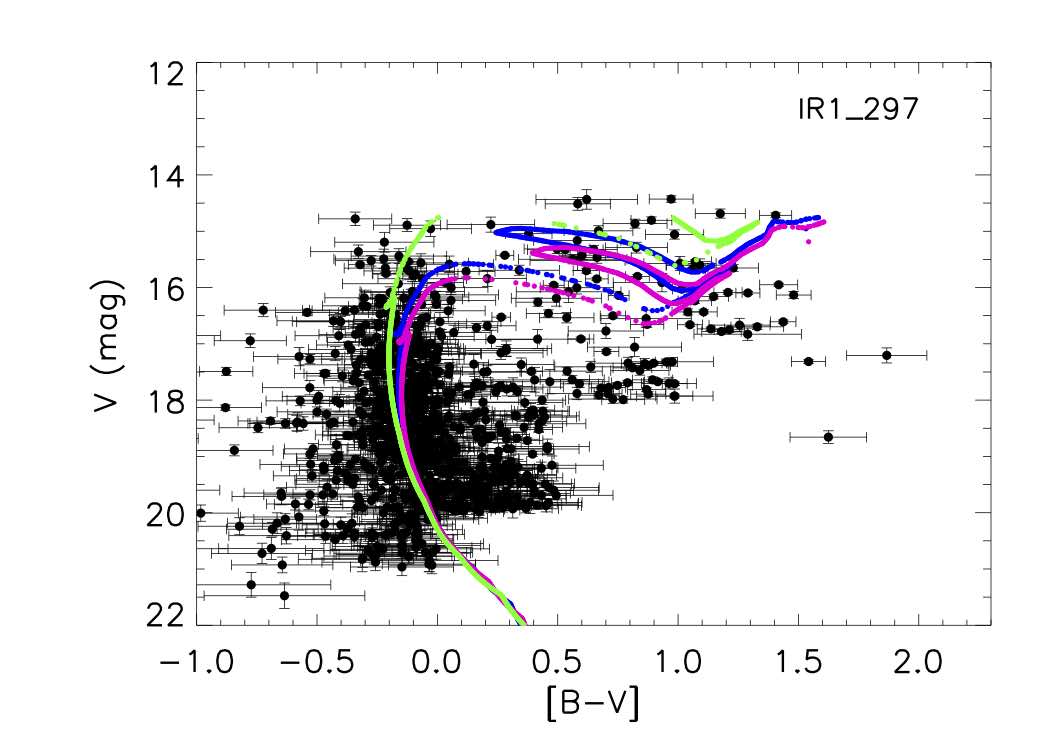}
\caption{An example of the $V-i$ versus $i$ CMD fitting: cluster candidate IR1-297. {\it Blue:} isochrone with the highest likelihood that corresponds to 117 Myr age; {\it green:} isochrone in the 16$^{\rm th}$ percentile (71 Myr); {\it magenta:} isochrone in the 84$^{\rm th}$ percentile (130 Myr). Only stars with $p_{\rm memb}>$ 0.60 are considered. }
\label{fig_isoc}
\end{center}
\end{figure}

\subsection{The cluster age determination algorithm}

To determine the ages of our clusters we used a Bayesian approach to obtain the most likely theoretical isochrone that reproduces the observed CMD. The method is analytically described in \citet{Hernandez2008}, and \citet{Walmswell13}. In this method, each theoretical isochrone is treated as a probability density function (PDF) on the CMD. Provided that, the probability of a cluster star $i$ located at ($x_{i}\pm\sigma_{x,i}$, $y_{i}\pm\sigma_{y,i}$) to come from a model isochrone $n$ is: 
\begin{equation}
p_{i}=\int\rho_{n}(x,y)U_{i}(x-x_{i},y-y_{i})dxdy
\end{equation}
where $U_{i}(x-x_{i},y-y_{i})$ is the error function for the star $i$, and $\rho_{n}(x,y)$ is the PDF along the model isochrone $n$. The error function of the star is taken as a bivariate Gaussian:
\begin{equation}
U_{i}(x-x_{i},y-y_{i})=\frac{1}{2\pi\sigma_{x,i}\sigma_{y,i}}e^{-\left(\frac{(x-x_{i})^{2}}{2\sigma_{x,i}^{2}}+\frac{(y-y_{i})^{2}}{2\sigma_{y,i}^{2}}\right)},
\end{equation}
where the star $i$ is associated with errors $\sigma_{x,i}$ and $\sigma_{y,i}$. Finally, the total likelihood of the cluster to be drawn from the model isochrone $n$ is given by the product of probabilities over the $S$ stars belonging to the cluster:
\begin{equation}
L_{n}=\prod_{i=1}^{S}p_{i}
\end{equation}

The complete set of such likelihoods ($L_{n}$) provides the PDF of the age of the given cluster. The assumed cluster age is the maximum of the PDF, with lower and upper uncertainties calculated as the 16$^{\rm th}$ and 84$^{\rm th}$ percentiles, respectively (e.g., see Fig.~\ref{fig_isoc}). The code we use was originally developed to infer the star formation histories of systems of few observed stars (such as ultra faint dwarf galaxies and star clusters), and is analytically described in Ram\'irez-Siordia et al. (in prep.), where multiple comparisons with real and simulated clusters are performed to test its accuracy, as well as the effects on estimated ages from the uncertainties on metallicity, distance modulus and extinction. A difference with the original code is that our version uses $p_{\rm memb}$, described in Section 4.1, for the estimation of equation (2) ($p_{i}'=p_{i}\cdot p_{memb}$). Therefore, the final likelihood of the star to belong to the isochrone will be given by the product of the two probabilities. Moreover, the code is allowed to vary the distance modulus by $\pm$0.25 mag (this corresponds to $\approx$11 kpc), in order to account for distance variations of the clusters in the LMC \citep{Subramanian09}. To ensure a robust age estimation, we fit the ($U-V$) versus $V$, ($B-V$) versus $V$, and ($V-i$) versus $i$ CMDs of each cluster, and we combine the results as described in the next section. A comparison between the  ages derived from the three different CMDs, yields median ratios of 1.0 to 1.1 with a standard deviation of the ratios $\sim$5.5. 

The isochrones we used are a byproduct of an independent project by Charlot \& Bruzual (in preparation)\footnote{The Charlot \& Bruzual isochrones are available to the interested user upon request.}. These authors have assembled complete sets of PARSEC evolutionary tracks computed by \citet{Chen15} for 16 values of the stellar metallicity, ranging from $Z = 0.0001$ to $Z = 0.06$, complemented with the work by \citet{Marigo13} to follow the evolution of stars through the thermally pulsing asymptotic giant branch (TP-AGB) phase. The isochrone synthesis algorithm described by \citet{CB91} is used to build isochrones from the evolutionary tracks at any age. Whereas the galaxy spectral evolution models by Charlot \& Bruzual use a large number of empirical and theoretical stellar libraries to describe the spectrophotometric properties of the stars along these isochrones \citep[e.g.,][]{Wofford16, Gutkin16, Vidal17}, for the purpose of this investigation we use the BaSeL 3.1 atlas \citep{Westera02} to obtain the stellar $UBVi$ magnitudes. The effects of dust shells surrounding TP-AGB stars on their spectral energy distribution is treated as in \citet{RAGL10}. The BaSeL 3.1 atlas covers uniformly the $(T_{eff}, log\ g)$ plane at the LMC metallicity of $Z$=0.008 ($[Fe/H]= -0.34$), according to studies of Cepheid LMC stars \citep[e.g.,][]{Romaniello05, Keller06}. Clusters with ages $\sim$2-3 Gyr are expected to have lower metallicities \citep[e.g. $Z$=0.006;][]{Piatti15}, but this difference is insignificant compared to the uncertainties in the estimation of their age. Our final grid of 80 isochrones covers the range 6.9 $\le$ log(age) $<$ 9.7 yr. 

%The isochrones we used were created especially for our project using the stellar evolutionary tracks from \citet{Charlot17}; they have a fixed metallicity of $Z$=0.008 (or $[Fe/H]= -0.34$ dex), based on the results of the studies of Cepheid LMC stars \citep[e.g.][]{Romaniello05, Keller06}. However, clusters with ages $\sim$2-3 Gyr are expected to have lower metallicities \citep[e.g. $Z$=0.006;][]{Piatti15}, yet that difference is insignificant compared to the uncertainties in the estimation of their age. Finally, the age grid we used covers the range 6.9 $\ge$ log(age) $>$ 9.7 yr, for a total of 80 isochrones.

\begin{figure*}
\begin{center}
\includegraphics[scale=0.4]{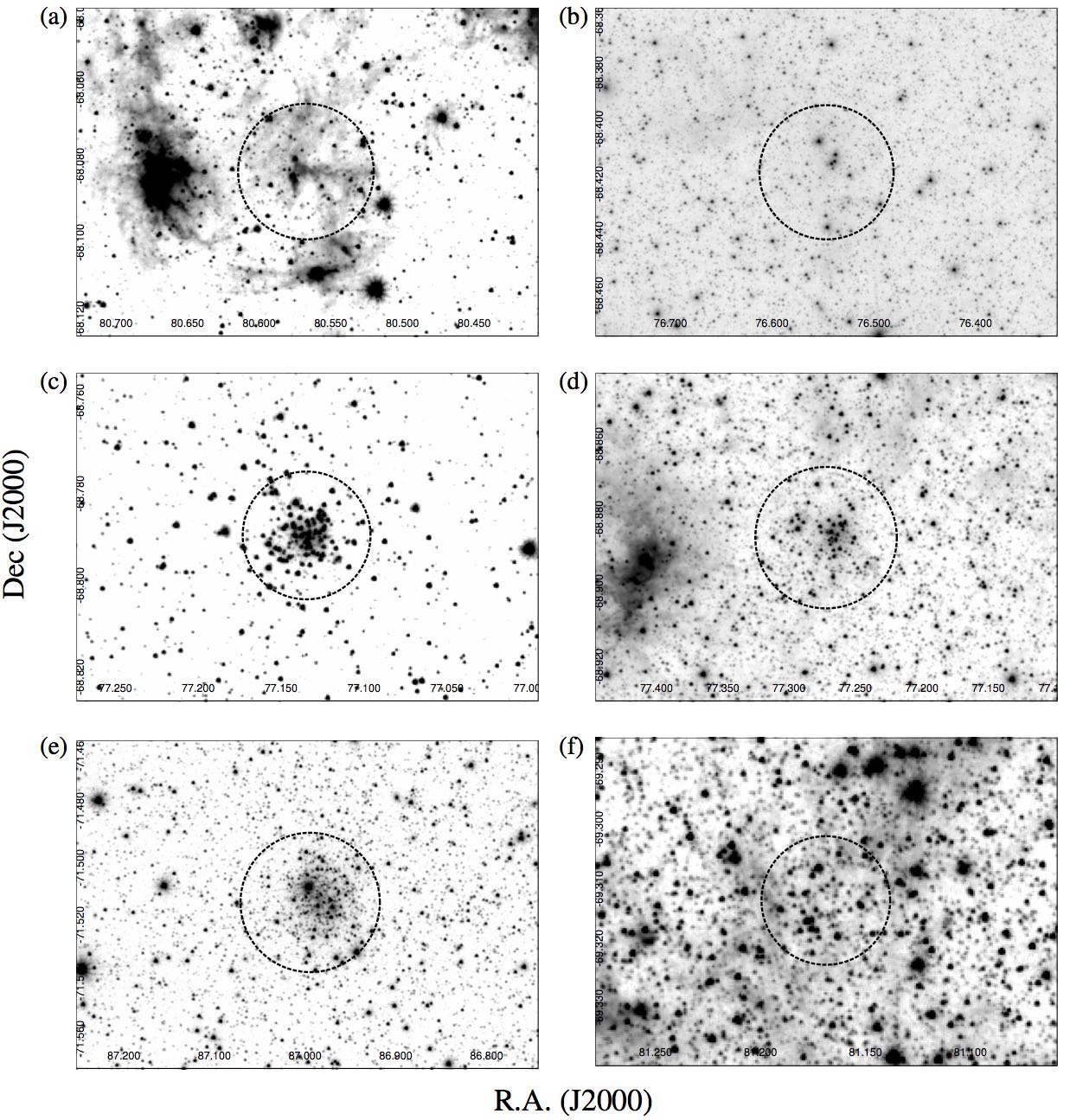}
\caption{Examples of clusters from our catalog presented on the Spitzer $IRAC~3.6\micron$ image. The dashed black lines mark the radii, as defined by the star-counts code. (a): cluster NUV-729, age 12.7$^{+9.2}_{-3.6}$ Myr; (b) NUV-1711, age 18.2$^{+2.7}_{-0.4}$ Myr; (c) IR1-1718, age 53.7$^{+10.8}_{-6.9}$ Myr; (d) IR1-1700, age 81.2$^{+9.9}_{-7.1}$ Myr; (e) IR1-284, age 549$^{+226}_{-113}$ Myr; and (f) IR1-1169, age 4.89$^{+0.72}_{-2.26}$ Gyr. Note that clusters (a) and (e) were not included in the catalog compiled by \citet{Bica08}.}
\label{fig_clusters}
\end{center}
\end{figure*}

\subsection{Final star cluster catalog}
Our final catalog contains 4850 clusters and is presented in Table~\ref{tab_clusters}. Column (1) gives the cluster ID assigned by our detection code. Columns (2) and (3), respectively, report the right ascension (R.A.) and declination (Dec.) of the cluster centers, in J2000 decimal equatorial coordinates. The cluster radii are presented in column (4). Finally, columns (5), (6), and (7) contain the best age estimation for each cluster, as well as its lower and upper uncertainty bounds (from the 16$^{\rm th}$ and 84$^{\rm th}$ percentiles of the PDF). 

The best age for each cluster is calculated as a combination of the ages resulting from the fitting of the aforementioned three CMDs [i.e., ($U-V$), ($B-V$), and ($V-i$)], after weighting them by both the number of stars that were used in each fit and the total uncertainty of the corresponding age measurement. This is synopsized by the following formula: 
\begin{equation}
{\rm Best\ Age} = \frac{ \sum\limits_{i=1}^3 \frac{N_{i}\cdot {\rm Age}_{i}}{(\delta {\rm Age}_{i})} }{\sum \limits_{i=1}^3 \frac{N_{i} }{(\delta {\rm Age}_{i})} },
\end{equation}
where {\it i}=1-3 correspond to the results of the ($U-V$), ($B-V$), and ($V-i$) CMD fits, respectively. $N_{i}$ is the number of stars fitted in the {\it i-}th CMD,  {\rm Age}$_{i}$ is the best age estimation in that CMD, and $\delta${\rm Age}$_{i}$ is the difference between the upper and lower uncertainties of {\rm Age}$_{i}$. In Fig.~\ref{fig_clusters}, we present some characteristic examples of clusters from our catalog, ordered by increasing age. Although most of them were detected in the same band in which they are displayed (i.e. the Spitzer $IRAC~3.6\micron$), clusters (a) and (b) were only found in the UV bands, thus stressing the importance of using multiple bands in the identification procedure. 

The cross-correlation of our new cluster catalog with those of \citet{Bica08} and \citet{Werchan11} shows that we identify 3451 new clusters in the 7 deg$^{2}$ we surveyed. 

\begin{figure}
\begin{center}
\includegraphics[scale=0.5]{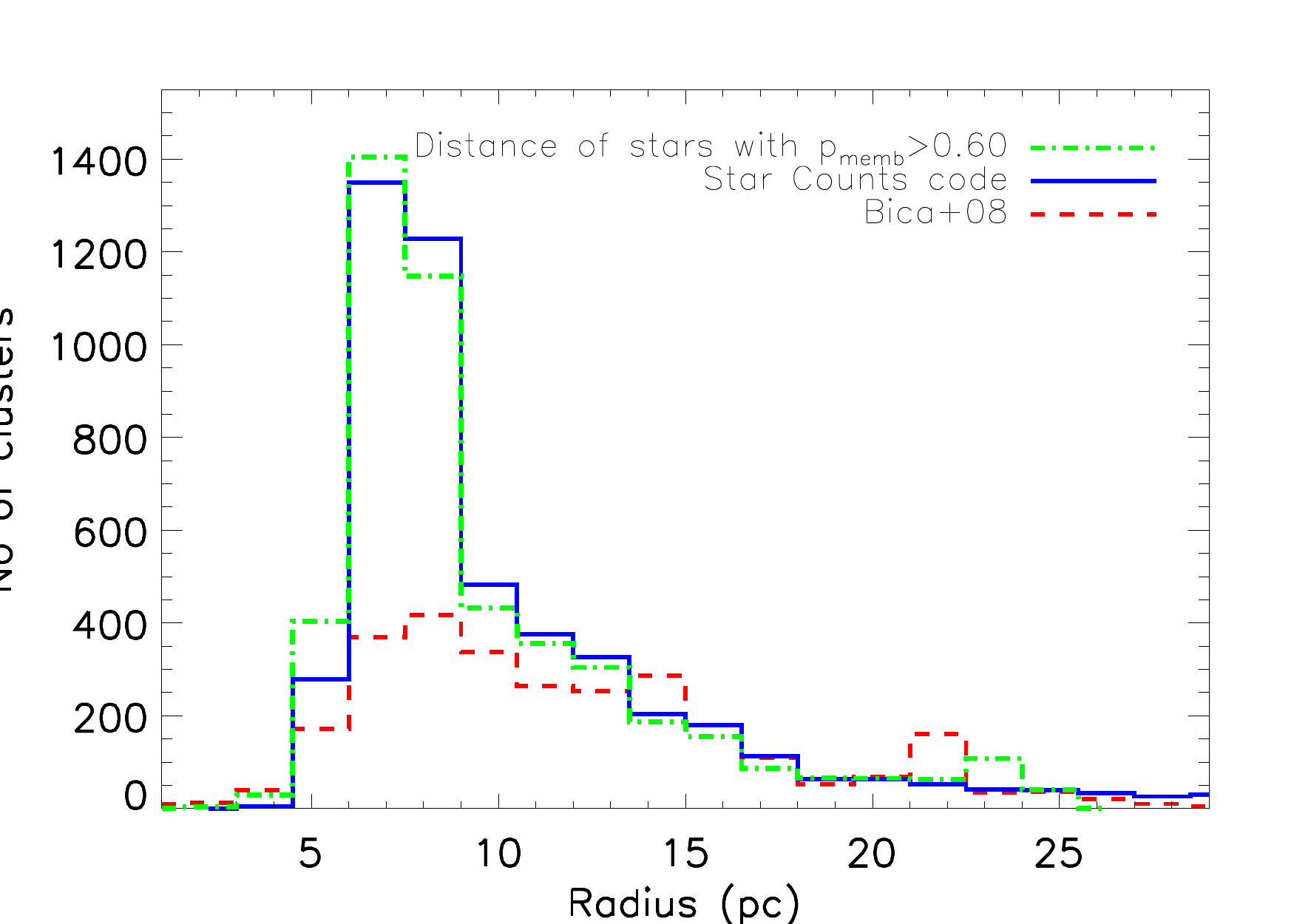}
\caption{The distribution of star cluster radii in our sample (solid blue line), compared with that from \citet[][dashed red line]{Bica08}. The dashed-dotted green line is also our distribution of radii, albeit derived in an independent way, using the furthest cluster star with $p_{\rm memb}>$ 0.60. }
\label{fig_radii}
\end{center}
\end{figure}

\begin{figure*}
\begin{center}
\includegraphics[scale=0.7]{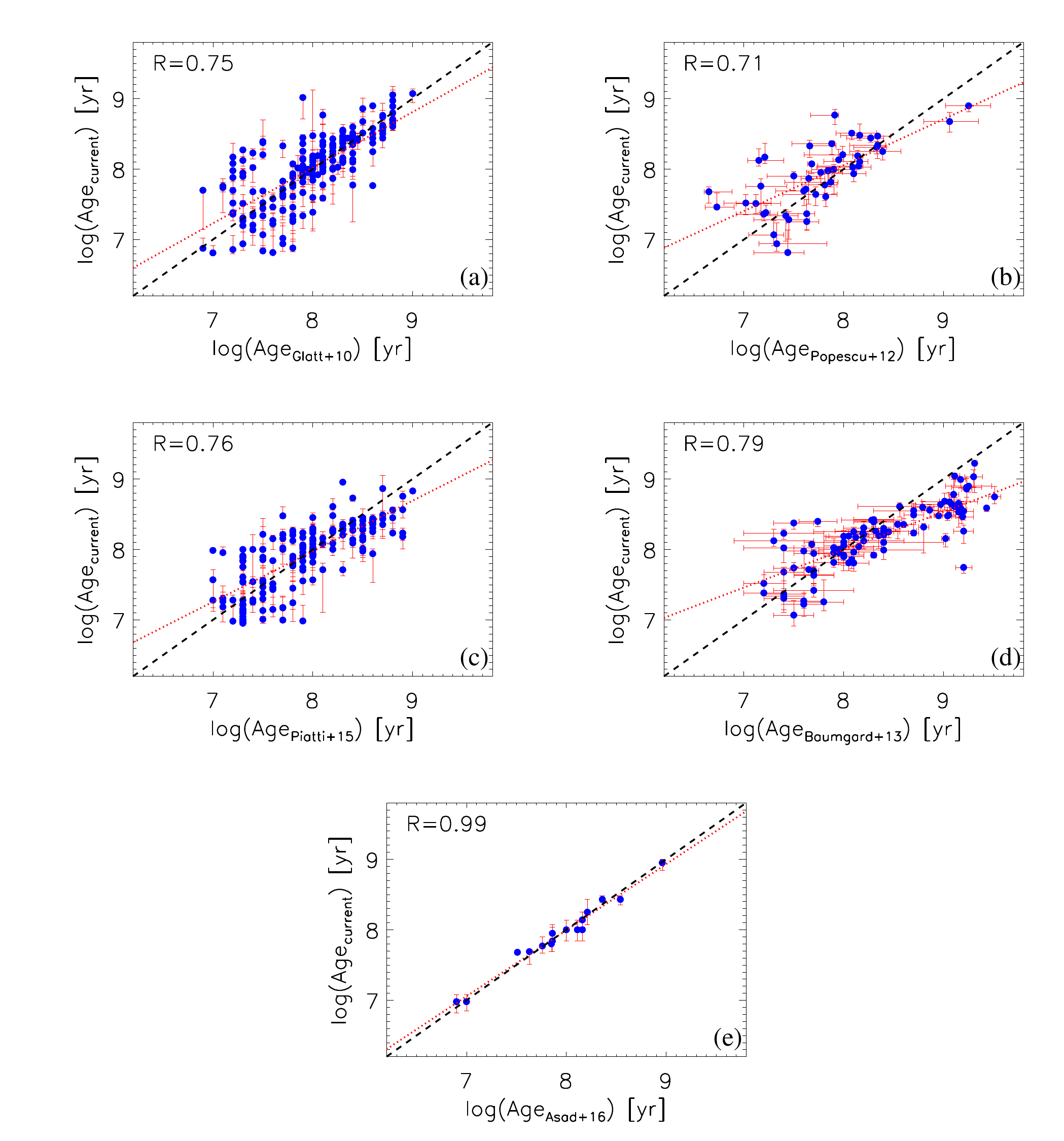}
\caption{Comparison of the ages determined from our method ($\rm Age_{\rm current}$) for clusters we have in common with (a) \citet{Glatt10}, (b) \citet{Popescu12}, (c) \citet{Piatti15}, (d) \citet{Baumgardt13}, and (e) \citet{Asad16}. The dashed black lines correspond to the one-to-one correlation, while the dotted red ones are the least square fits to the data. The Pearson correlation coefficients (R) are indicated in the upper left corner of each panel.}
\label{fig_comp_ages}
\end{center}
\end{figure*}

% ################################################
\section{Results}

\subsection{Comparisons of cluster radii with other surveys}

\begin{table*}
\begin{minipage}{120mm}
\begin{center}
\caption{Star cluster catalog.}
\label{tab_clusters}
\begin{tabular}{lcccccc}
\hline 
\hline
 & R.A.(J2000) & Dec(J2000) & Radius & log(Age) & Lower unc. & Upper unc. \\
 CID   & (deg)         & (deg)            & (deg)    & (yr) & (yr) & (yr) \\
\hline
  IR1-1     &  76.6915 &  -72.3672 &  0.0086  &   8.92 &  8.83 &  8.95  \\
  M2-1065 &   78.8299 &  -68.7034  &  0.0094   & 8.02  & 7.63  & 8.12  \\
  NUV-10  &    75.0914 &  -71.5908 &  0.0097  &   8.14 &  8.01 &  8.22 \\
  ... & ... & ... & ... & ... & ... & ... \\
\hline
\end{tabular}
\end{center}
{\bf Notes. } \\
The lower and upper uncertainty bounds are calculated as the 16$^{\rm th}$ and 84$^{\rm th}$ percentiles, respectively. (The full version is available online.)\\ 
\end{minipage}
\end{table*}% 

In Fig.~\ref{fig_radii} we compare the distribution of the radii of our clusters (estimated the Gaussian full width at half maximum of each cluster's counts), shown with a solid blue line, with that of the the semimajor axes of the clusters from the sample of \citet{Bica08}, presented with a dashed red line. Although they do not represent the exact same parameter, both distributions peak around 5-10 pc, with our maximum being more pronounced. Our method fails to detect clusters with radii smaller than 4 pc, a range in which \citet{Bica08} found 3\% of their clusters. At the distance of the LMC, 4 pc correspond to only 8 arcseconds. Since the spatial resolution of the images we used for the cluster identification varies from 2 to 5 arcseconds (0.5-1.25 pc), clusters with $r \leq 4$ pc will be hard to identify unambiguously. To ensure our code is not subject to some selection bias, we estimate the radii of our clusters in an independent way. In particular, we use the results of the membership probability (presented in Section 4.1) to find, for each cluster, the distance of the furthest cluster star with $p_{\rm memb}>$ 0.60, and then we adopt this as an alternative cluster radius. We over-plot these values in Fig.~\ref{fig_radii}, with a dotted-dashed green line and find that it is almost identical to our initial radii estimation. Another important difference with \citet{Bica08} is the absence of the minor peak at $\sim$21-22 pc. Such associations could originate after the expulsion of gas during  cluster formation, which usually results in rapid mass loss and eventually their dissolution \citep[e.g.][]{Pfalzner09}. Hence, the detection of these objects is very improbable due to their short lifetimes and low luminosities. Alternatively, this might be also a selection bias introduced by having set an upper limit (i.e., the region-of-interest) on the cluster radii during the detection process.

\begin{figure}
\begin{center}
\includegraphics[scale=0.5]{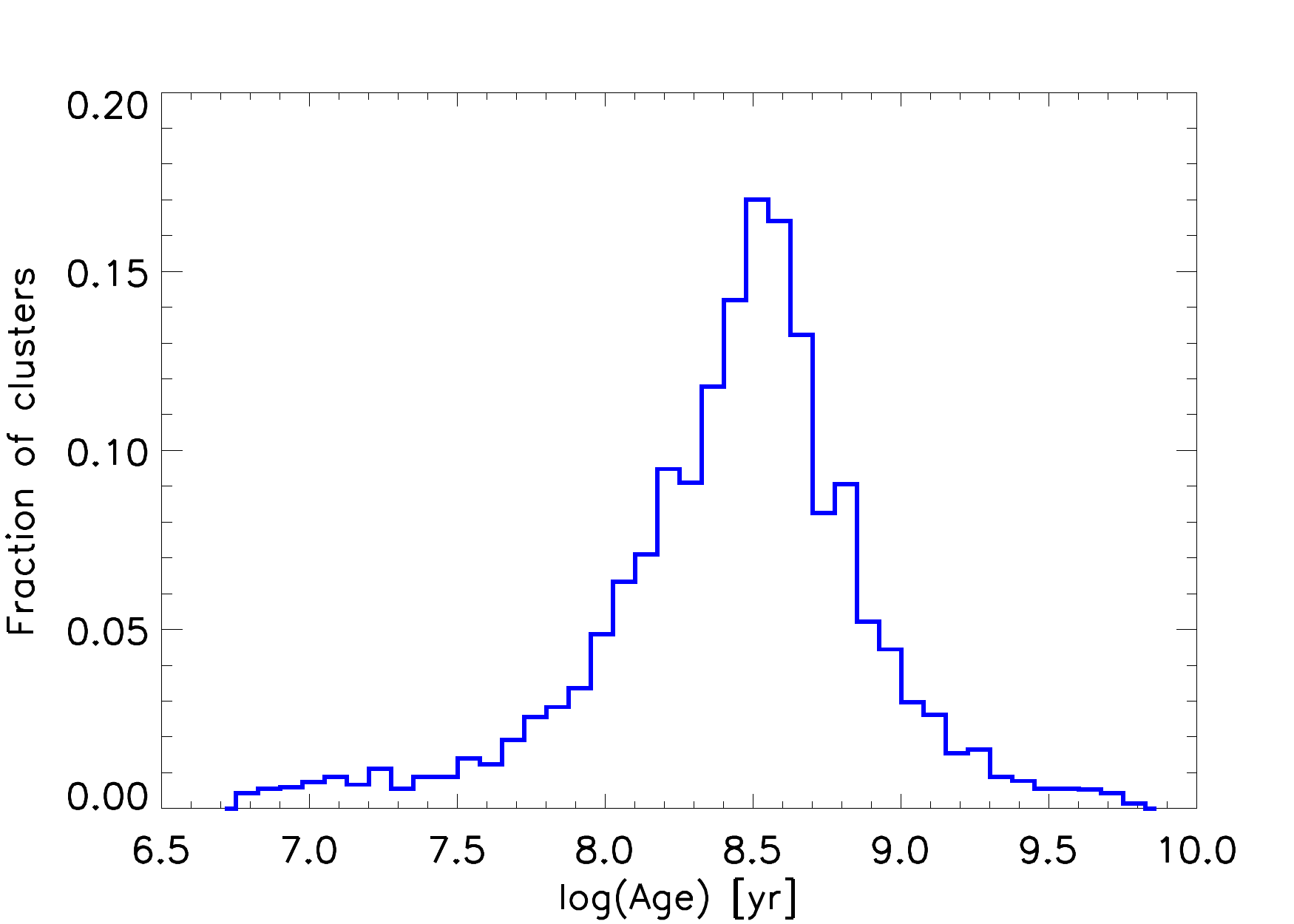}
\caption{Age distribution of our LMC star clusters. }
\label{fig_hist_ages}
\end{center}
\end{figure}

\subsection{Comparisons of cluster ages with other surveys}

In Fig.~\ref{fig_comp_ages} we compare our age estimations for clusters in common with other surveys. \citet{Glatt10} estimated ages by visually identifying the clusters' MSTOs. They also attempted to decontaminate their CMDs, again by visual means. In fact, the visual identification may underestimate the ages of some clusters, since the existence of blue stragglers or field stars projected on the extrapolation of the main sequence to brighter magnitudes might confuse the observer. Only a statistical method that fits the integrated CMD can overcome such limitations. There is nonetheless, as one can see in panel (a) of Fig.~\ref{fig_comp_ages}, a good correlation between the two works (with  Pearson R=0.75). In panel (b) of the same figure, we compare our age estimations with those from \citet{Popescu12}, who derived the ages for their clusters by comparing their observed colors with those from Monte-Carlo simulations. Although their method is more sophisticated than visual estimations, unfortunately it fails to remove the contamination of the field stars and therefore does not appear to correlate with our results better than the previously considered work by Glatt et al.; it yields R=0.71.  \citet{Piatti15} used high resolution VISTA Magellanic Cloud (VMC) survey near-infrared CMDs to estimate the ages of $\sim$ 300 clusters in the LMC bar and the 30 Dor region. They visually fitted theoretical isochrones on the ($Y-K$) versus $K$ CMDs, using a sophisticated field star decontamination technique \citep{Piatti14} that takes into account density variations in the field. Comparing their age estimations with ours yields R=0.76. \citet{Baumgardt13} selected the best age estimates from previous publications to compile their sample. Their results correlate slightly better with our measurements, having R=0.79. Finally, \citet{Asad16} estimated the ages of 27 massive LMC clusters by fitting their integrated cluster spectra with theoretical models \citep{Bruzual03}. Although their sample is very small, their results are in excellent agreement with ours, with R=0.99.

\subsection{Age distribution of the LMC star clusters }

\begin{figure*}
\begin{center}
\includegraphics[scale=0.63]{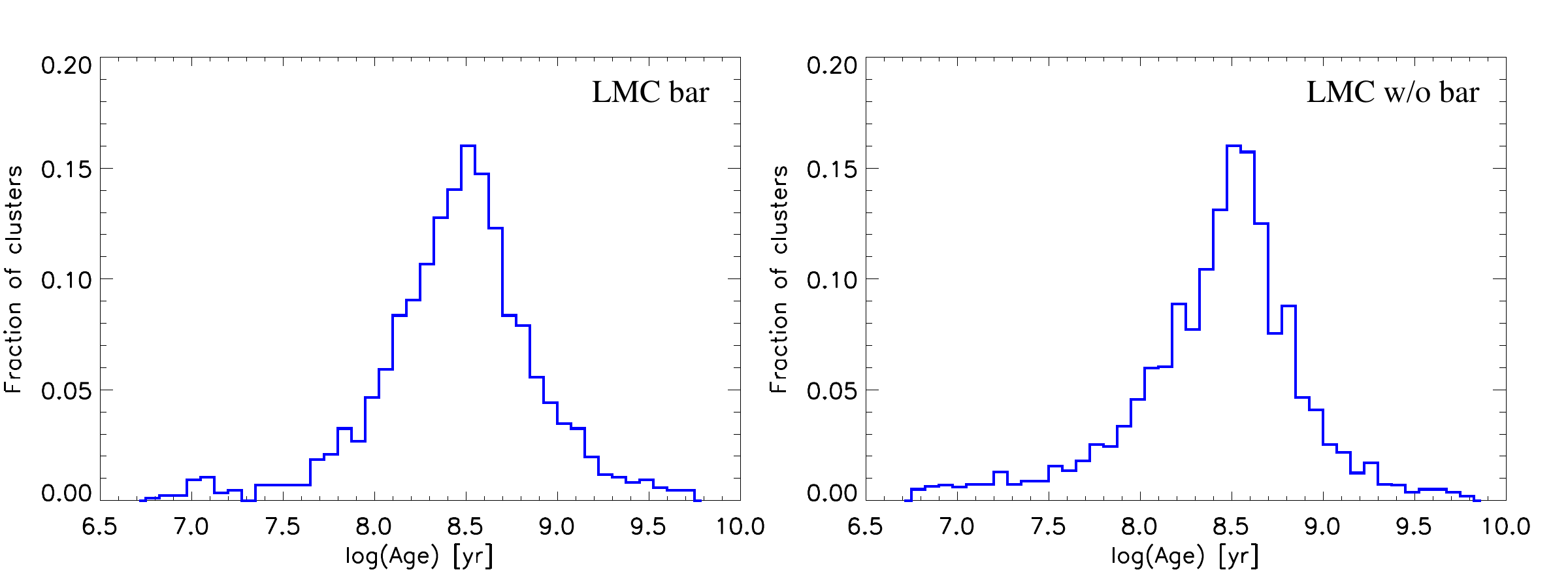}
\caption{Age distribution of the star clusters found in the LMC bar (left panel) and in the rest of the galaxy (right panel).}
\label{fig_hist_ages_bar}
\end{center}
\end{figure*}

\begin{figure*}
\begin{center}
\includegraphics[scale=0.7]{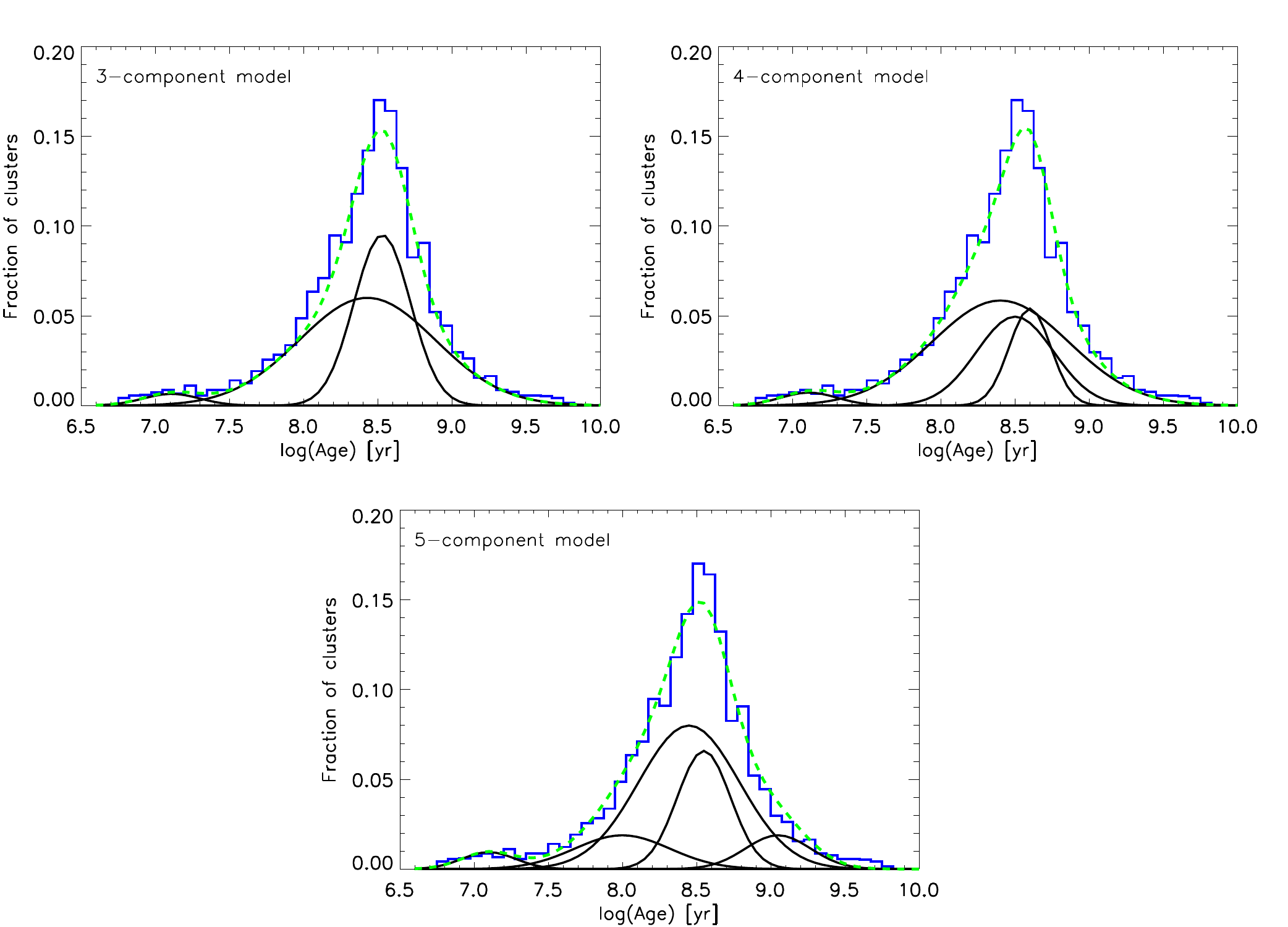}
\caption{The three (top left), four (top right), and five (bottom) component mixture models (dashed green lines), and their individual constituents (solid black lines). }
\label{fig_mix}
\end{center}
\end{figure*}

\begin{figure*}
\begin{center}
\includegraphics[scale=0.4]{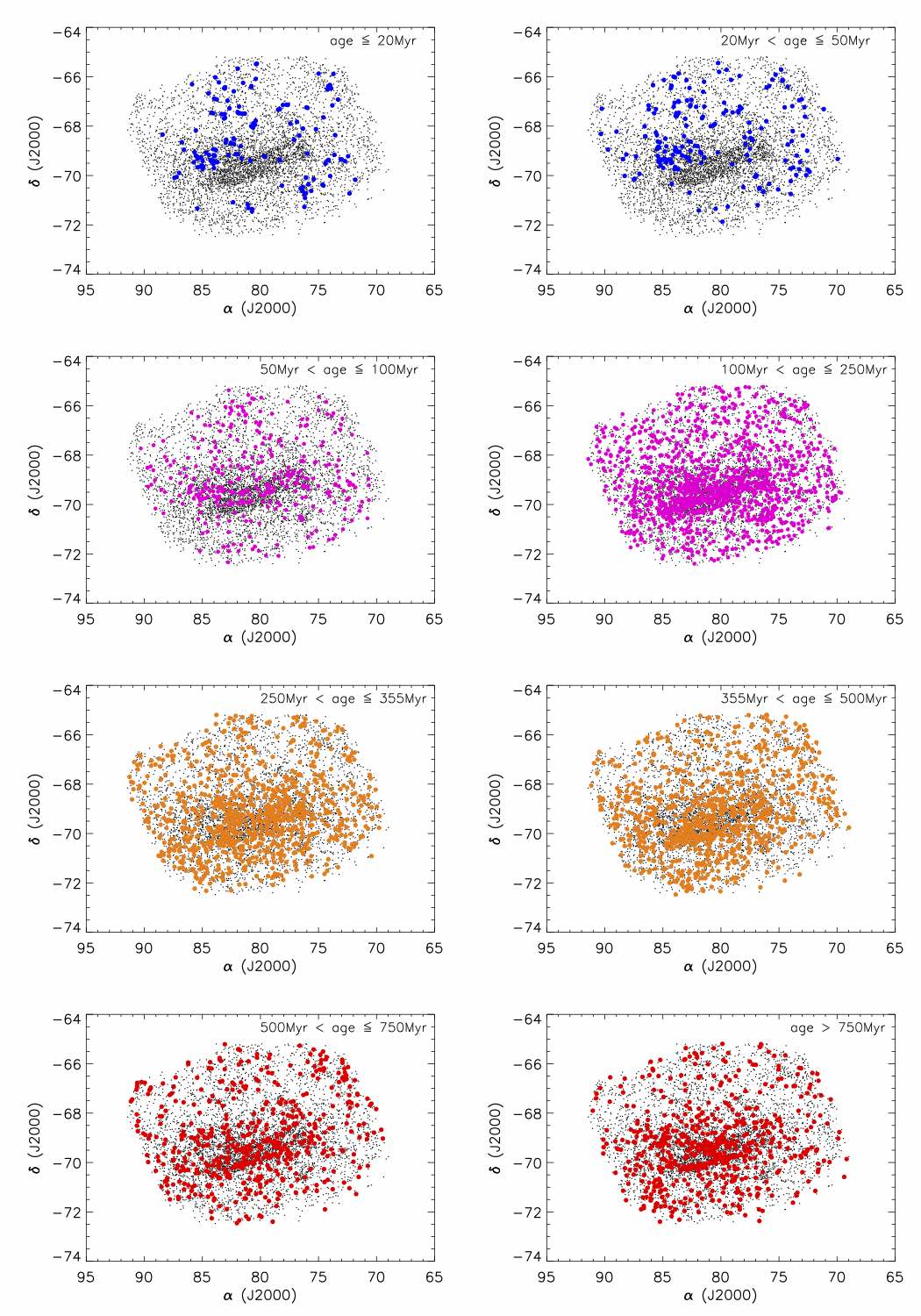}
\caption{Spatial distribution of ages for all the star clusters in our sample (in black dots). From top-left to bottom-right we present the positions of star clusters with:  {\rm Age}$\le$20 Myr, 20$<${\rm Age}$\le$50 Myr, 50$<${\rm Age}$\le$100 Myr, 100$<${\rm Age}$\le$250 Myr, 250$<${\rm Age}$\le$355 Myr, 355$<${\rm Age}$\le$500 Myr, 500$<${\rm Age}$\le$750 Myr, and {\rm Age}$>$750 Myr. }
\label{fig_sfh}
\end{center}
\end{figure*}

With cluster ages and positions available, one can identify prominent peaks in their formation history. This can provide important information about the LMC-SMC-Galaxy interactions in space and time. In Fig.~\ref{fig_hist_ages}, we present the age distribution of our clusters. Since histogram peaks can result as artifacts of the binning scheme, we adopted our bin size using the un-biased Freedman-Diaconis rule; its value is 0.075 dex (well above the resolution of our isochrones, which is 0.03 dex). The histogram shows the existence of at least one significant peak $\approx$310 Myr ago, with secondary ones 160 and 500 Myr ago. Moreover, a smaller increase appears at $\approx$10-20 Myr. Such enhancements in the cluster formation can be associated with an LMC-SMC direct collision (about 100-300 Myr ago) predicted by the models of \citet{Besla12}, or tidal interactions between the Clouds and the Galaxy. According to \citet{Besla07}, the LMC is still in its first passage about the Galaxy, and the related tides might have triggered star formation during the past Gyr. Similarly, \citet{Harris09} studied the star formation history of the LMC, by comparing the MCPS CMDs with theoretical models; they suggested various peaks of star formation (at 12, 100, 500 Myr, and 2 Gyr), many of which are also consistent with our analysis.

Interestingly, the LMC bar (selected loosely by us to lie between R.A. 76$^{\rm h}$ -- 85$^{\rm h}$ and Dec. -70.5$^{\rm o}$ -- -69.0$^{\rm o}$; see left panel of Fig.~\ref{fig_hist_ages_bar}) harbors fewer peaks than anywhere else in the galaxy (right panel of the same figure): there, cluster formation appears only at $\sim$10 and 310 Myr ago.  

Moreover, only 50\% of the clusters we detect are older than $\approx$300 Myr. Although it is true that the MSTOs of such clusters are well below our detection limit, their scarcity is probably real, owing to cluster dissolution. Star clusters are destroyed through various mechanisms, such as residual gas expulsion \citep[e.g][]{Baumgardt07}, two-body relaxation, and external tidal fields \citep[e.g.,][]{Baumgardt03}, as well as via tidal heating from shocks, and the harassment from the passage of giant molecular clouds \citep[e.g.,][]{Gnedin97,Gieles06}. According to \citet{Baumgardt13}, as many as 90\% of the clusters with ages greater than 200 Myr can be dissolved per dex of lifetime: this could explain the decline of the cluster population older than $\sim$500 Myr in our sample.

Albeit cluster dissolution might explain why the age distribution resembles a Gaussian, one can see that additional populations might be drawn by fitting its exact shape using multiple components. For that purpose we used a code for Bayesian analysis of univariate Gaussian mixtures (NMIX\footnote{Publicly available at \url{https://people.maths.bris.ac.uk/~mapjg/Nmix}}). This code implements the approach presented in \citet{Richardson97}. The results suggest that the above distribution can be fitted either by three, four or five components (see Fig.~\ref{fig_mix}); these solutions have Bayes K-factors between them of $\sim$1, while those including more or fewer components fail to provide good fits (when comparing the successful models to each one of the rejected univariate distributions, we get K-factors $>$ 5). In Fig.~\ref{fig_mix}, we show the three successful mixture models (in dashed green lines,) as well as their individual components (solid black lines). It is worth noticing here that the use of variable width in normal (Gaussian) components is a reasonable approximation, since it is expected that the age uncertainties will dominate over the length of the individual star cluster formation events. According to the results of NMIX, the LMC has experienced various epochs of cluster formation over the past Gyr, with the most probable 10, 310, and 400 Myr ago; these events seem to agree with the visually identified peaks, presented above. Additionally,  possible secondary peaks exist 70 Myr, 250 Myr, and 1.1 Gyr ago.

Important information can also be drawn from the spatial distribution of the cluster ages. In Fig.~\ref{fig_sfh} we present that distribution (in decimal equatorial coordinates), with ages separated into 8 different bins of varying sizes. As shown, younger clusters ($\le$50 Myr) are mostly distributed in the northern and northeastern regions of the LMC  (where the arm is located), as well as around its bar; they tend to gather in the regions indicated by \citet{Kim03} to host HI supershells (see figure 6 of that article), and in the star-forming region 30 Dor. On the other hand, older clusters (250-500 Myr) appear very concentrated in the LMC bar, and relatively uniformly distributed outside of that. These results are consistent with an inside-out formation scenario for the LMC --- at least for the past 1 Gyr. \citet{Cioni09} concluded likewise, based on the observed flattening of the metallicity gradient in the LMC. She argued that, as the galaxy builds-up inside-out, star formation moves towards the outer parts of the galaxy with the result that those regions  become enriched by metals and the metallicity gradient flattens out \citep[see also][and references therein]{Vlajic09}.

%Finally, the star cluster formation history at $>$500 Myr is mostly concentrated in the south-eastern region of the galaxy.   

The information presented above can be used not only to study the integrated star formation history of the LMC, but also to constrain dynamical simulations of the interactions among the LMC, the SMC, and the Galaxy. It is also important to infer the properties of the clusters, their relation with the environment, and how they might compare to clusters in other galaxies. In a forthcoming paper, we will apply our novel technique to study the cluster populations of the SMC and other nearby galaxies.

% ################################################
\section{Conclusions}
We present here a new, fully-automated method we have developed to detect and estimate the ages of star clusters in nearby galaxies with resolved stellar populations. The detection  is performed using a simple, but robust, algorithm (complemented by Monte-Carlo test simulations) to find overdensities even in the most crowded fields. Our procedure decontaminates the cluster candidate color-magnitude diagrams from field stars and, using a Bayesian isochrone fitting code, estimates their ages. We apply our method on multi-band, high-resolution archival images of the Large Magellanic Cloud and we conclude the following.
\begin{enumerate}

\item[(a)] Out of 4850 clusters in the 7 deg$^{2}$ we surveyed, 3451 have not been reported before. Of those, $\sim$150 (3\%) have ages $\le$20 Myr. Young clusters can contribute unique information for determining the stellar initial mass function.

\item[(b)] The distribution of cluster radii is consistent with the expected sizes of clusters in the LMC, peaking at 5-10 pc.

\item[(c)] The results on the age distribution of our star clusters are consistent with various epochs of star cluster formation in the LMC. The most prominent occurred $\sim$310 ago, with secondary ones 10, 160, and 500 Myr ago. All these episodes could be the result of interactions between the LMC, the SMC, and the Galaxy, as suggested by the findings of N-body simulations \citep[e.g.,][]{Yoshizawa03, Besla12}. We also show that the age distribution of the LMC bar star clusters is different from that of the rest of the galaxy. 

\item[(d)] The spatial distribution of the clusters as a function of age suggests that the youngest clusters are located outside the LMC bar (mostly in the northern and northeastern regions of the galaxy), whereas clusters older than 750 Myr preferentially sit in the bar. The regions with the highest concentration of young clusters ($<$50 Myr) are those identified by \citet{Kim03} to host HI supershells. The above results might suggest an inside-out star cluster formation scenario for the LMC, during the past 1 Gyr.

\end{enumerate}

\acknowledgments

The authors would like to thank Dr. A. Piatti for insightful comments that helped improve this paper, and Dr. H. Oti for helping with the data acquisition in the initial stages of this investigation. The authors wish to thank the anonymous referee for her/his thorough review and valuable comments. TB would like to acknowledge support from the CONACyT Research Fellowships program. We gratefully acknowledge support from the program for basic research of CONACyT through grant number 252364. GM acknowledges support from CONICYT, Programa de Astronom\'ia/PCI, FONDO ALMA 2014, Proyecto No 31140024, and GA\,\v{C}R under grant number 14-21373S. RAGL thanks DGAPA, UNAM, for support through the PASPA program. GB acknowledges support for this work from UNAM through grant PAPIIT IG100115. This research made use of TOPCAT, an interactive graphical viewer and editor for tabular data. IRAF is distributed by the National Optical Astronomy Observatory, which is operated by the Association of Universities for Research in Astronomy (AURA) under cooperative agreement with the National Science Foundation.

\bibliography{LMCbib.bib} % if your bibtex file is called example.bib

%\end{thebibliography}



\end{document}